\documentclass[aps,prd,preprint,nofootinbib,superscriptaddress]{revtex4-2}

\usepackage{amsmath,amssymb,bm}
\usepackage{graphicx}
\usepackage{booktabs}
\usepackage{siunitx}
\usepackage{hyperref}
\usepackage{microtype}
\usepackage{enumitem}
\hypersetup{
  colorlinks=true,
  linkcolor=blue,
  citecolor=blue,
  urlcolor=blue,
  pdfauthor={Gazal Sharma and Gaurav Katoch},
  pdftitle={Minimal Quark--Lepton Complementarity from a Rigid Deformation of Tri-Bimaximal Mixing}
}

\newcommand{\UTBM}{U_{\rm TBM}}
\newcommand{\VCKM}{V_{\rm CKM}}
\newcommand{\UPMNS}{U_{\rm PMNS}}
\newcommand{\Vgg}{V_c^{gg}}

\newcommand{\dcp}{\delta_{\rm CP}}
\newcommand{\Jcp}{J_{\rm CP}^{\ell}}
\newcommand{\Ggg}{\Gamma_{gg}}
\newcommand{\Sgg}{\Sigma_{gg}}
\newcommand{\STBM}{\Sigma_{\rm TBM}}

\begin{document}

\title{Minimal Quark--Lepton Complementarity from a Rigid Deformation of Tri-Bimaximal Mixing}

\author{Gazal Sharma}
\email[Corresponding author: ]{gazal.sharma555@gmail.com}
\affiliation{Centre for Research Impact and Outcome, Chitkara University, Rajpura, Punjab, India}
\affiliation{Department of Physics, Sharda School of Engineering \& Science, Sharda University, Greater Noida, India}
\author{Gaurav Katoch}
\email{gauravkatoch1988@gmail.com}
\affiliation{Department of Physics, Rayat Bahra University, Mohali, Punjab, India}

\date{20 August 2026}

\begin{abstract}
We propose a minimal three-family realization of quark--lepton complementarity (QLC) in which the non-trivial correlation matrix is an exactly unitary, rigid deformation of tri-bimaximal mixing. In a canonical TBM convention the deformation is a rotation about the normalized axis $n_{gg}=(2,1,2)^T/3$, with rotation angle fixed by the Cabibbo angle, $\alpha_{gg}=-3\theta_C/4$. The resulting ansatz,
$\UPMNS=\VCKM^\dagger\UTBM R_{n_{gg}}(-3\theta_C/4)$,
contains no new continuous parameter in the Dirac lepton-mixing sector once the structural choice is fixed. With the 2025 Particle Data Group CKM fit it predicts
$\theta_{12}=33.29^\circ$, $\theta_{13}=8.46^\circ$, $\theta_{23}=49.19^\circ$, and $\dcp=180.79^\circ$, with CKM-induced uncertainties much smaller than present oscillation errors. We derive leading Wolfenstein sum rules; in particular, the real deformation preserves the CKM-induced suppression $\Jcp=-A\eta\lambda^3/6+\mathcal O(\lambda^4)$ and hence a near-CP-conserving phase. A retrospective simplicity scan shows that the pair $(2,1,2),3/4$ ranks first among 5510 oriented primitive-integer/rational candidates when tested on epoch-matched 2016 and 2018 oscillation data. In an enlarged 49,622-candidate 2018 scan it is second in raw score but remains first among candidates no more complex than itself under several independent simplicity measures; profiling the rotation strength for the fixed $(2,1,2)$ axis gives $r_{\rm BF}=0.74941$. We formulate the deformation by a convention-covariant flavor-space generator and give an exact effective eigenframe realization $M_\nu=A I+B\Sgg+C\Sgg^2$. The construction is presently allowed in normal ordering, with the atmospheric sector providing its strongest pressure; its upper-octant atmospheric angle and near-$\pi$ Dirac phase provide sharp falsifiability targets.
\end{abstract}

\maketitle

\section{Introduction}

The observed disparity between quark and lepton mixing remains one of the central structural puzzles of flavor physics. The CKM matrix is close to the identity, whereas the PMNS matrix contains two large mixing angles and a sizable reactor angle. No organizing principle for this pattern is part of the Standard Model, and symmetry-based attempts to understand fermion masses and mixings remain an active field of study \cite{Ding:2024vyn}.

Quark--lepton complementarity (QLC) was motivated by the approximate relation between the Cabibbo and solar angles, $\theta_C+\theta_{12}^{\ell}\simeq\pi/4$ \cite{Minakata:2004xt}, and has been explored in unified and matrix-based formulations \cite{Antusch:2005ca}. A particularly useful formulation is to regard the product of quark and lepton mixing matrices as a non-trivial correlation matrix. In that language, early phenomenological studies showed that a correlation matrix with a small or vanishing $(1,3)$ entry could predict $\theta_{13}^{\rm PMNS}$ near $9^\circ$ before the reactor angle was precisely measured \cite{Chauhan:2006im}. A later numerical reconstruction of the QLC correlation matrix found a texture significantly closer to tri-bimaximal (TBM) than to bi-maximal mixing and used it to constrain the atmospheric angle and CP invariants \cite{Sharma:2015rva}.

Exact TBM mixing \cite{Harrison:2002er} is excluded by the non-zero reactor angle, but it remains a useful zeroth-order eigenframe for flavor constructions. Cabibbo-sized perturbations of lepton mixing (``Cabibbo haze''), unitary perturbations about TBM, and symmetry-based QLC departures from TBM were developed well before the present work \cite{Datta:2005ci,Pakvasa:2007xv,Ahn:2011yj}. Likewise, additional $R_{23}$ or $R_{13}$ rotations of the TBM neutrino basis and charged-lepton rotations have been studied systematically \cite{Shimizu:2014tia}; SO(3) family symmetry with Pati--Salam unification can relate charged-lepton corrections to quark mixing \cite{King:2005bj}; and tri-bimaximal-Cabibbo mixing directly tied $\theta_{13}$ to $\theta_C$ \cite{King:2012in}. Exponential/axis-angle descriptions of lepton mixing and alternative rotation-based formulations of QLC also exist \cite{Zhukovsky:2019qlc}. More recently, CKM-multiplied BM, TBM and golden-ratio structures, with additional perturbations introduced to fit the complete lepton data, have been analyzed in detail \cite{Giarnetti:2024qlc}, while the first JUNO results have already begun to discriminate among minimally modified TBM patterns \cite{He:2026tbm}.

Our purpose is therefore not to claim novelty for QLC, TBM, Cabibbo corrections, rotation parametrizations, or quark-related charged-lepton corrections separately. The object tested here is narrower: a \emph{single rigid} correlation-matrix deformation whose canonical TBM-axis representation is $(2,1,2)/3$, whose angle is locked to $-3\theta_C/4$, and which consequently fixes all four Dirac PMNS observables once the CKM matrix is supplied. To our knowledge, after comparison with QLC correlation-matrix, TBM-deformation, CKM--PMNS factorization and exponential/rotation parametrizations, we have not identified an earlier construction equivalent to this specific rigid $gg$ realization. This statement is intentionally narrower than a claim of mathematical uniqueness, since a generic three-dimensional rotation can always be Euler-decomposed into products of plane rotations.

The answer developed here is the matrix $\Vgg$. In a fixed canonical TBM convention it is specified by an oriented axis $n_{gg}=(2,1,2)^T/3$ and a rotation angle $-3\theta_C/4$. The strict model has no new continuous parameter in the Dirac lepton-mixing sector. We emphasize from the outset that this is a phenomenological quark--lepton complementarity/flavor-correlation construction, not a model of gauge quark--lepton unification and not yet a complete ultraviolet derivation of the axis or of the factor $3/4$. Accordingly, a substantial part of this work is devoted to robustness: exact unitarity, analytic sum rules, a historical simplicity scan, a convention audit, and present-data falsifiability.

The paper is organized as follows. Section~\ref{sec:model} defines the exact ansatz and its convention-covariant generator. Section~\ref{sec:analytic} derives its leading analytic consequences. Section~\ref{sec:numerics} gives present predictions and propagated CKM uncertainties. Section~\ref{sec:historical} presents the retrospective historical and simplicity tests. Section~\ref{sec:mass} gives an exact effective flavor-spurion realization. Section~\ref{sec:data} discusses the present oscillation status and falsifiability. We conclude in Sec.~\ref{sec:conclusion}. Technical details are collected in the appendices.

\section{The minimal $\Vgg$ ansatz}
\label{sec:model}

We work with the real TBM convention
\begin{equation}
\UTBM=
\begin{pmatrix}
\sqrt{\frac23} & \frac{1}{\sqrt3} & 0\\
-\frac{1}{\sqrt6} & \frac{1}{\sqrt3} & \frac{1}{\sqrt2}\\
\frac{1}{\sqrt6} & -\frac{1}{\sqrt3} & \frac{1}{\sqrt2}
\end{pmatrix}.
\label{eq:utbm}
\end{equation}
The canonical coordinate representation of the $gg$ rotation axis is
\begin{equation}
 n_{gg}=\frac{1}{3}(2,1,2)^T,
 \qquad n_{gg}^T n_{gg}=1.
\label{eq:ngg}
\end{equation}
For any normalized vector $n$, define the cross-product generator
\begin{equation}
[n]_\times=
\begin{pmatrix}
0&-n_3&n_2\\
n_3&0&-n_1\\
-n_2&n_1&0
\end{pmatrix},
\end{equation}
and the exact SO(3) rotation
\begin{equation}
R_n(\alpha)=\exp\!\left(\alpha[n]_\times\right)
=I+\sin\alpha\,[n]_\times+(1-\cos\alpha)[n]_\times^2.
\label{eq:rodrigues}
\end{equation}

The proposed QLC correlation matrix is
\begin{equation}
\boxed{\Vgg=\UTBM R_{n_{gg}}(\alpha_{gg})},
\label{eq:vcgg}
\end{equation}
with the strict Cabibbo locking
\begin{equation}
\boxed{\alpha_{gg}=-\frac34\theta_C,\qquad \theta_C\equiv\arcsin\lambda.}
\label{eq:alpha}
\end{equation}
The PMNS matrix is then
\begin{equation}
\boxed{
\UPMNS=\VCKM^\dagger\Vgg
=\VCKM^\dagger\UTBM R_{n_{gg}}\!\left(-\frac34\theta_C\right).
}
\label{eq:master}
\end{equation}
Equation~\eqref{eq:master} is the defining equation of the minimal model. It corresponds to the minimal phase-aligned version of matrix QLC: additional diagonal quark--lepton mismatch phases, sometimes introduced in more general correlation-matrix formulations, are set to unity. Majorana phases may be multiplied on the right of $\UPMNS$ and do not affect the oscillation predictions discussed here.

For later comparison it is useful to define the one-parameter extension
\begin{equation}
\mathcal M_{gg}(r):\qquad
\UPMNS(r)=\VCKM^\dagger\UTBM R_{n_{gg}}(-r\theta_C),
\label{eq:rmodel}
\end{equation}
for which the strict QLC point is $r=3/4$.

\subsection{Convention-covariant flavor-space form}

The coordinate triple $(2,1,2)$ depends on the signs chosen for the TBM eigenvectors and should not by itself be treated as a basis-independent observable. The convention-independent content with respect to TBM column-sign rephasings is carried by the flavor-space generator
\begin{equation}
\Ggg\equiv\UTBM[n_{gg}]_\times\UTBM^T=[n_f]_\times,
\qquad n_f\equiv\UTBM n_{gg}.
\label{eq:gamma}
\end{equation}
Explicitly,
\begin{equation}
 n_f=
\begin{pmatrix}
\dfrac{1+2\sqrt2}{3\sqrt3}\\[4pt]
\dfrac{1-\sqrt2+\sqrt6}{3\sqrt3}\\[4pt]
\dfrac{-1+\sqrt2+\sqrt6}{3\sqrt3}
\end{pmatrix}
\simeq
\begin{pmatrix}
0.7368\\0.3917\\0.5511
\end{pmatrix}.
\label{eq:nf}
\end{equation}
The equivalent flavor-space rotation is
\begin{equation}
Q_{gg}(\alpha)=e^{\alpha\Ggg}
=\UTBM R_{n_{gg}}(\alpha)\UTBM^T,
\label{eq:qgg}
\end{equation}
so that
\begin{equation}
\UPMNS=\VCKM^\dagger Q_{gg}\UTBM.
\label{eq:mastergamma}
\end{equation}
Under a TBM column-sign transformation $\UTBM\to\UTBM S$, with $S={\rm diag}(\pm1,\pm1,\pm1)$, the coordinates of the axis transform simultaneously so that $\Ggg$ is unchanged. Under a general weak-basis transformation $W$, $\Ggg$ transforms covariantly as $\Ggg\to W\Ggg W^T$. Thus $\Ggg$, rather than the bare coordinate triple, is the appropriate object for an eventual flavor-theory embedding.

\section{Analytic consequences}
\label{sec:analytic}

We use the standard Wolfenstein organization \cite{Wolfenstein:1983yz},
\begin{equation}
 s_{12}^q=\lambda,\qquad s_{23}^q=A\lambda^2,
 \qquad s_{13}^qe^{i\delta_q}=A\lambda^3(\rho+i\eta),
\end{equation}
with Eq.~\eqref{eq:master} evaluated as
\begin{equation}
\UPMNS=\VCKM^\dagger\UTBM
R_{n_{gg}}\!\left[-\frac34\arcsin\lambda\right].
\end{equation}
To first order in $\lambda$, the three mixing angles are
\begin{align}
\theta_{12}
&=\theta_{12}^{\rm TBM}
+\frac{1-\sqrt2}{2}\lambda+\mathcal O(\lambda^2),
\label{eq:t12exp}\\
\theta_{13}
&=\left(
\frac{\sqrt2}{2}+\frac{\sqrt6}{12}-\frac{\sqrt3}{6}
\right)\lambda+\mathcal O(\lambda^2),
\label{eq:t13exp}\\
\theta_{23}
&=\frac{\pi}{4}
+\frac{\sqrt3+2\sqrt6}{12}\lambda+\mathcal O(\lambda^2).
\label{eq:t23exp}
\end{align}
The reactor coefficient is
\begin{equation}
\frac{\sqrt2}{2}+\frac{\sqrt6}{12}-\frac{\sqrt3}{6}
\simeq0.6226,
\end{equation}
while the positive coefficient in Eq.~\eqref{eq:t23exp} shows that the upper-octant tendency is structural rather than an accidental numerical feature.

Eliminating $\lambda$ at leading order gives two simple lepton-only sum rules,
\begin{align}
\theta_{12}-\theta_{12}^{\rm TBM}
&\simeq-0.3327\,\theta_{13},
\label{eq:sum12}\\
\theta_{23}-45^\circ
&\simeq0.8876\,\theta_{13},
\label{eq:sum23}
\end{align}
where the same angular unit is understood on both sides.

Because the $gg$ deformation is real, all leptonic Dirac CP violation originates from the CKM phase in the minimal construction. The Jarlskog invariant \cite{Jarlskog:1985ht} has the leading form
\begin{equation}
\boxed{
\Jcp=-\frac{A\eta}{6}\lambda^3+\mathcal O(\lambda^4).
}
\label{eq:jcp}
\end{equation}
The same leading coefficient occurs in the exact CKM--TBM parent limit studied in Ref.~\cite{Giarnetti:2024qlc}; it is therefore not a novelty claim of the present construction. Rather, the $gg$ deformation preserves the characteristic CKM-induced $\mathcal O(\lambda^3)$ suppression while fixing a different set of angle correlations. Since $J_{\rm CP}^{\rm max}\propto s_{13}=\mathcal O(\lambda)$, Eq.~\eqref{eq:jcp} implies
\begin{equation}
\dcp=\pi+\mathcal O(\lambda^2),
\end{equation}
i.e. near-CP conservation is a parametric prediction of the real model.

A further exact property follows because the rotation leaves its axis invariant:
\begin{equation}
\Vgg\,(2,1,2)^T=\UTBM(2,1,2)^T.
\label{eq:eigenframe}
\end{equation}
Equation~\eqref{eq:eigenframe} is best interpreted as an eigenframe relation in the chosen canonical convention; the equivalent flavor-space statement is encoded by Eqs.~\eqref{eq:gamma}--\eqref{eq:qgg}.

\section{Numerical predictions from present CKM data}
\label{sec:numerics}

For the present numerical evaluation we use the 2025 Particle Data Group global CKM fit \cite{PDG2025},
\begin{equation}
\lambda=0.22501\pm0.00068,\quad
A=0.826^{+0.016}_{-0.015},\quad
\bar\rho=0.1591\pm0.0094,\quad
\bar\eta=0.3523^{+0.0073}_{-0.0071}.
\label{eq:pdginputs}
\end{equation}
Using the exact standard CKM parametrization and the exact rotation in Eq.~\eqref{eq:master}, the strict model gives
\begin{equation}
\boxed{
\theta_{12}=33.2899^\circ,\quad
\theta_{13}=8.4554^\circ,\quad
\theta_{23}=49.1909^\circ,\quad
\dcp=180.7905^\circ.
}
\label{eq:predictions}
\end{equation}
The corresponding Jarlskog invariant is
\begin{equation}
\Jcp\simeq-4.50\times10^{-4}.
\end{equation}

A Monte Carlo propagation of the current CKM errors gives the uncertainties listed in Table~\ref{tab:pred}. In this propagation the standard CKM parameters are sampled independently; hence the table should be read as a transparent input-error estimate rather than a covariance-complete global uncertainty analysis.

\begin{table}[t]
\caption{Frozen $\Vgg$ predictions using the 2025 PDG CKM fit. The quoted errors are propagated CKM-input uncertainties from an independent-parameter Monte Carlo.}
\label{tab:pred}
\begin{ruledtabular}
\begin{tabular}{lcc}
Observable & Prediction & CKM-input error\\
\hline
$\theta_{12}$ & $33.289^\circ$ & $0.009^\circ$\\
$\theta_{13}$ & $8.455^\circ$ & $0.028^\circ$\\
$\theta_{23}$ & $49.188^\circ$ & $0.046^\circ$\\
$\dcp$ & $180.791^\circ$ & $0.021^\circ$\\
$\Jcp$ & $-4.51\times10^{-4}$ & $1.2\times10^{-5}$\\
\end{tabular}
\end{ruledtabular}
\end{table}

The small widths in Table~\ref{tab:pred} are important conceptually. Once the structural hypotheses are frozen, present CKM uncertainties do not provide enough freedom to move the PMNS predictions appreciably. Future disagreement in $\theta_{23}$ or $\dcp$ therefore cannot be absorbed by modest changes in the quark inputs.

\section{Retrospective historical and simplicity tests}
\label{sec:historical}

The axis $(2,1,2)$ and the Cabibbo coefficient $3/4$ were identified while examining modern PMNS information. Consequently, a fit to current data cannot be interpreted as an independent discovery significance. We therefore perform a retrospective robustness test using historical, epoch-matched quark and neutrino inputs. This is not a prospective blind test; rather, it asks whether the same simple structure would already have been selected by earlier data within a pre-defined class of simple alternatives.

\subsection{Discrete candidate class}

We scan
\begin{equation}
\UPMNS=\VCKM^\dagger\UTBM R_{\hat n}(-r\theta_C),
\qquad
\hat n=\frac{(a,b,c)}{\sqrt{a^2+b^2+c^2}},
\label{eq:scanmodel}
\end{equation}
where $(a,b,c)$ is an \emph{oriented} primitive integer triple and $r=p/q$ is a positive reduced rational number. Opposite axes are retained separately because, for a fixed negative rotation angle, $R_{-n}(-r\theta_C)=R_n(+r\theta_C)$.

The baseline scan uses
\begin{equation}
|a|,|b|,|c|\le3,\qquad 1\le p,q\le5,
\end{equation}
which gives 290 oriented primitive axes and 19 rational coefficients, or
\begin{equation}
\boxed{5510\ \text{candidate rotations}.}
\end{equation}
The enlarged scan uses $|a_i|\le5$ and $p,q\le8$, giving 1154 oriented axes and 43 rational coefficients,
\begin{equation}
\boxed{49\,622\ \text{candidates}.}
\end{equation}
For ranking we use a deliberately simple asymmetric-Gaussian consistency score constructed from the published one-dimensional $1\sigma$ intervals. It is not a substitute for the original multidimensional likelihood.

For 2016 we use the contemporaneous PDG CKM fit \cite{PDG2016} together with the well-determined $\theta_{12}$ and $\theta_{13}$ measurements; the atmospheric octant sensitivity was below $1\sigma$ at that time \cite{Esteban:2016qun}. For 2018 we use the corresponding PDG CKM fit \cite{PDG2018} and all three mixing angles from the global analysis without the additional Super-K atmospheric table \cite{Esteban:2018azc}. The central inputs used by the reproducibility code are collected in Table~\ref{tab:historicalinputs}; the asymmetric neutrino errors enter the ranking score exactly as described in Appendix~\ref{app:repro}.

\begin{table}[t]
\caption{Epoch-matched central inputs used in the retrospective scans. Mixing angles are in degrees. The 2016 $\theta_{23}$ entry is omitted from the score because the octant sensitivity was below $1\sigma$.}
\label{tab:historicalinputs}
\begin{ruledtabular}
\begin{tabular}{c@{\quad}cccc@{\qquad}ccc}
Year & $\lambda$ & $A$ & $\bar\rho$ & $\bar\eta$ & $\theta_{12}$ & $\theta_{13}$ & $\theta_{23}$\\
\hline
2016 & 0.22506 & 0.811 & 0.124 & 0.356 & 33.56 & 8.46 & --\\
2018 & 0.22453 & 0.836 & 0.122 & 0.355 & 33.82 & 8.61 & 49.6\\
\end{tabular}
\end{ruledtabular}
\end{table}

\begin{figure}[t]
\centering
\includegraphics[width=0.88\textwidth]{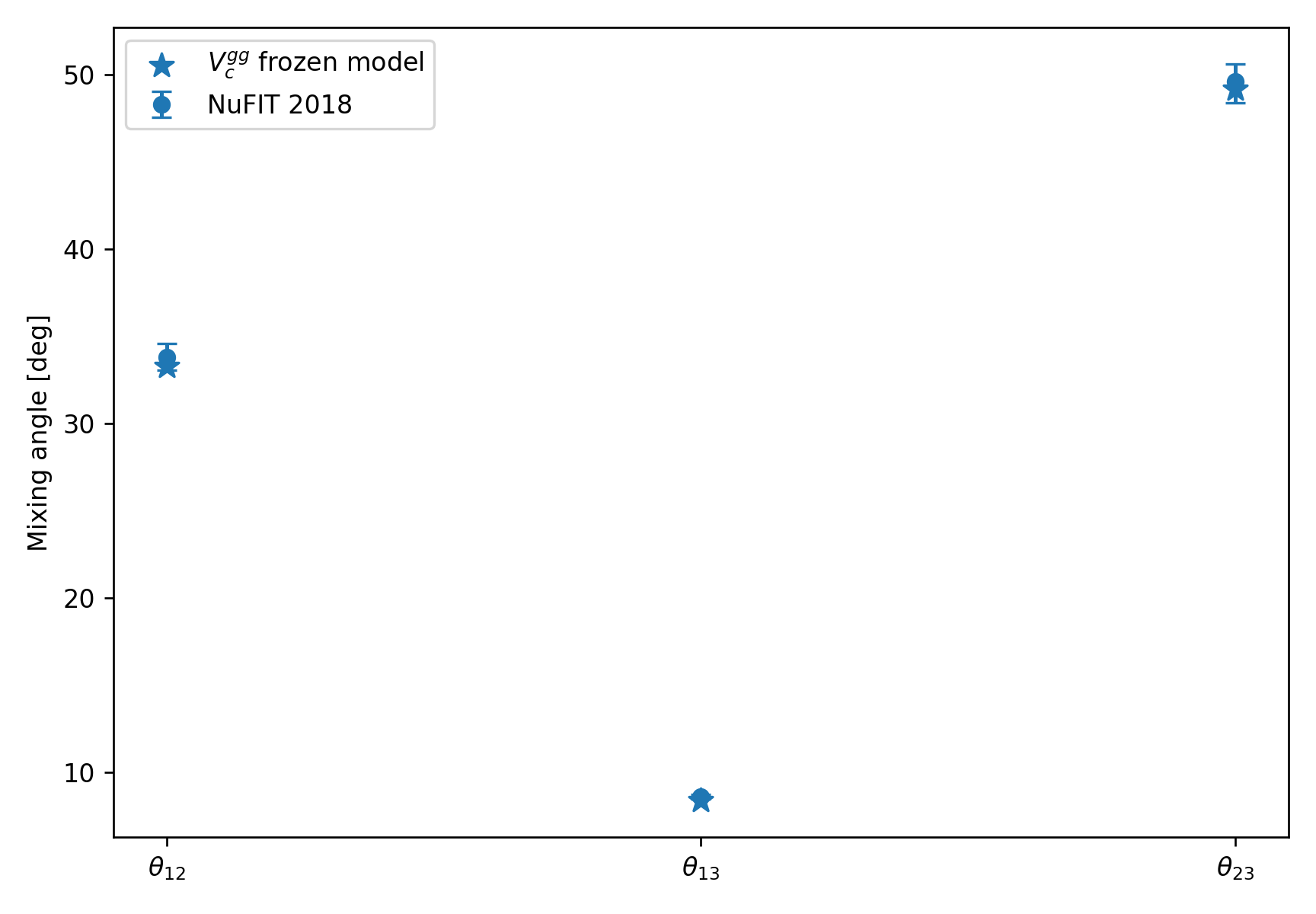}
\caption{Epoch-matched 2018 comparison between the frozen $V_c^{gg}$ prediction and the published normal-ordering global-fit values used in the retrospective scan. The figure is illustrative of compatibility only; correlations among the fitted oscillation parameters are not represented.}
\label{fig:historical2018}
\end{figure}

In the baseline 5510-candidate scan, $(2,1,2),3/4$ ranks first in both 2016 and 2018. The epoch-matched predictions and scores are
\begin{align}
2016:&\quad (33.282^\circ,\ 8.451^\circ),
\qquad \chi^2_{\rm score}=0.142,\\
2018:&\quad (33.274^\circ,\ 8.420^\circ,\ 49.161^\circ),
\qquad \chi^2_{\rm score}=2.777.
\end{align}
The historical 2018 comparison is shown in Fig.~\ref{fig:historical2018}. The point is frozen by the QLC prescription; the error bars are the one-dimensional 2018 global-fit intervals used only for this retrospective consistency test.

\begin{figure}[t]
\centering
\includegraphics[width=0.88\textwidth]{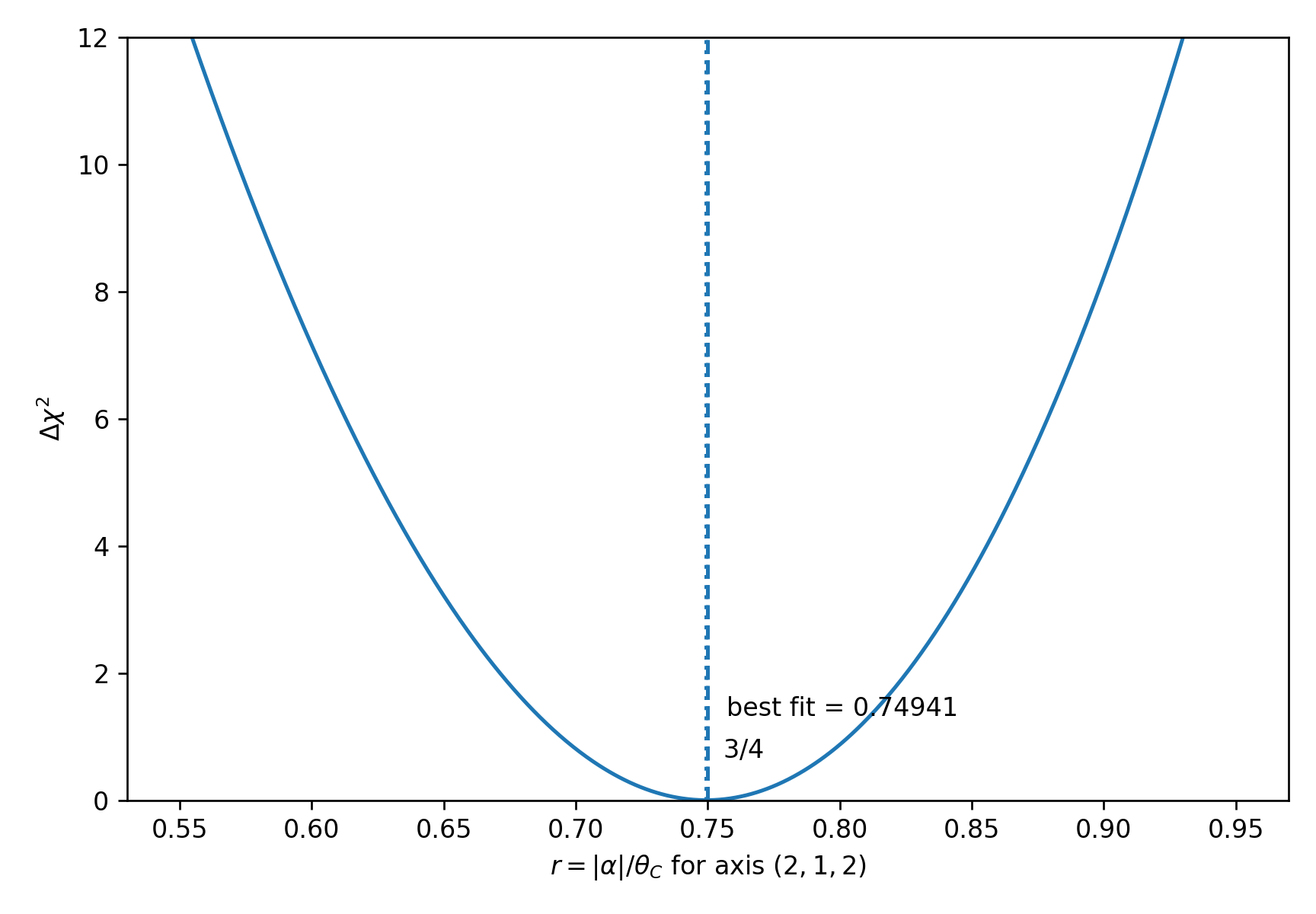}
\caption{Retrospective 2018 profile of the rotation strength $r=|\alpha|/\theta_C$ for the fixed canonical axis $(2,1,2)$. The curve is an asymmetric-Gaussian consistency score built from the published 2018 one-dimensional angle intervals, not the original full NuFIT likelihood. The continuous optimum $r=0.74941$ lies essentially on $3/4$.}
\label{fig:rprofile}
\end{figure}

The 2018 continuous profile in the fixed $(2,1,2)$ direction gives
\begin{equation}
\boxed{r_{\rm BF}^{2018}=0.74941,}
\label{eq:rbest}
\end{equation}
with
\begin{equation}
\Delta\chi^2_{\rm score}(r=3/4)\simeq1.15\times10^{-4}.
\end{equation}
Figure~\ref{fig:rprofile} displays this profile. The numerical proximity to $3/4$ should not be interpreted as evidence for a fundamental rational law: the profile is conditioned on the already selected $(2,1,2)$ direction and is retrospective. In the strict model $3/4$ remains a structural hypothesis whose value is supported here only by minimality and historical robustness, not dynamically derived.

\subsection{Simplicity versus raw fit quality}

\begin{figure}[t]
\centering
\includegraphics[width=0.88\textwidth]{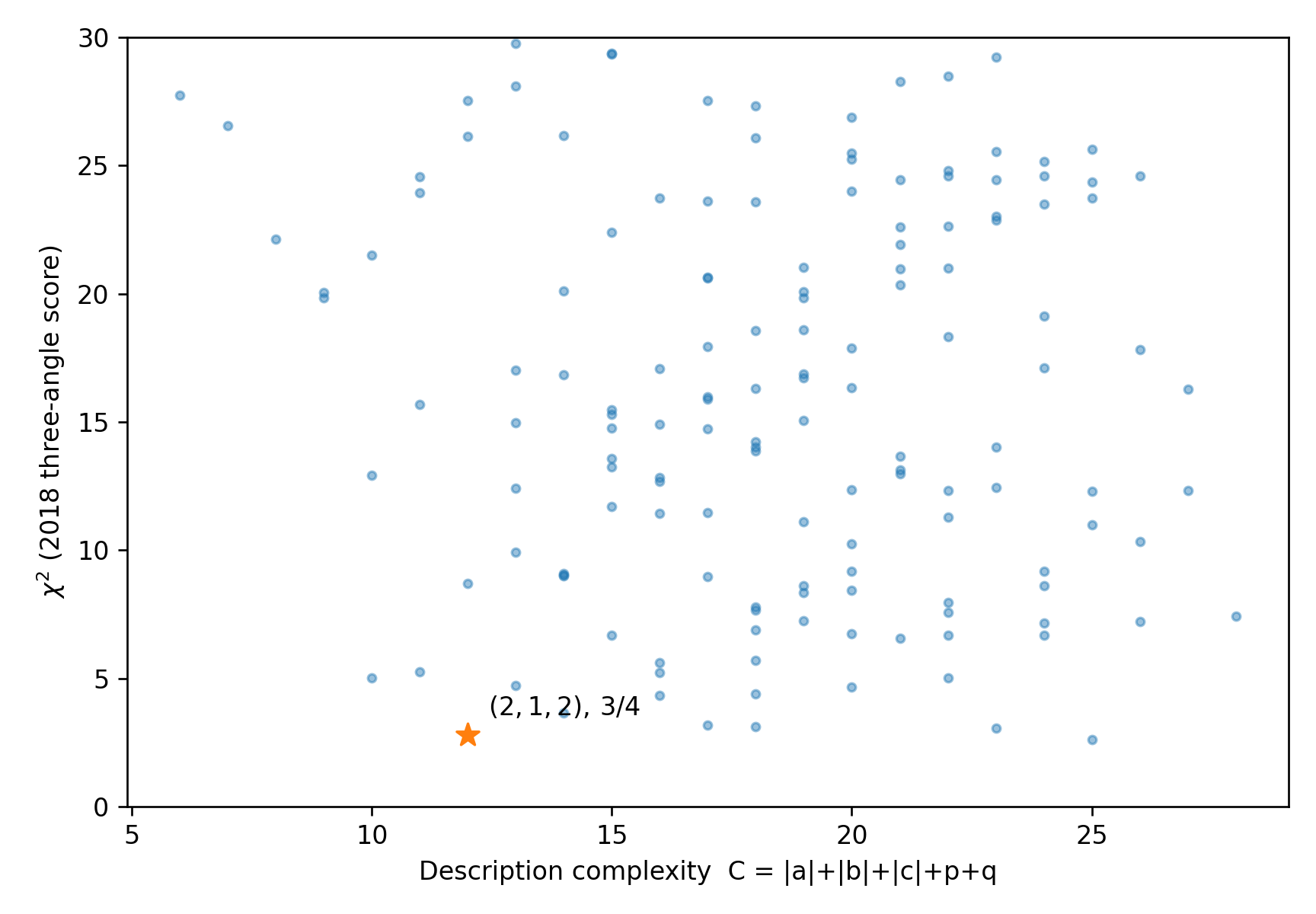}
\caption{Description complexity versus the 2018 three-angle consistency score for the enlarged discrete scan. The strict $(2,1,2),3/4$ point is the best tested candidate at or below its complexity $C=12$, though not the absolute raw-score minimum once substantially more complicated integer/rational structures are admitted.}
\label{fig:complexity}
\end{figure}

In the enlarged 49,622-candidate 2018 scan, the strict $gg$ model ranks second in raw score. The best raw-scoring point is a substantially more complicated choice, $(5,-4,3)$ with $r=8/5$. To quantify the tradeoff, we define a simple description complexity
\begin{equation}
C=|a|+|b|+|c|+p+q.
\label{eq:complexity}
\end{equation}
The strict model has $C_{gg}=12$, whereas the raw best point has $C=25$. No tested candidate with $C\le12$ has a lower 2016 or 2018 score than $(2,1,2),3/4$. The relevant 2018 comparison is shown in Table~\ref{tab:scan} and Fig.~\ref{fig:complexity}.

\begin{table}[t]
\caption{Selected points from the enlarged 2018 scan. The score is the historical asymmetric-Gaussian angle consistency statistic.}
\label{tab:scan}
\begin{ruledtabular}
\begin{tabular}{ccccc}
Axis $(a,b,c)$ & $r$ & $C$ & $\chi^2_{\rm score}$ & Raw rank\\
\hline
$(5,-4,3)$ & $8/5$ & 25 & 2.625 & 1\\
$(2,1,2)$ & $3/4$ & 12 & 2.777 & 2\\
$(4,2,5)$ & $5/7$ & 23 & 3.065 & 3\\
$(4,2,5)$ & $3/4$ & 18 & 3.114 & 4\\
$(2,1,2)$ & $5/7$ & 17 & 3.187 & 5\\
\end{tabular}
\end{ruledtabular}
\end{table}

This is the scientifically relevant outcome of the scan: the ansatz is not mathematically unique, but it is Pareto-efficient within the tested simple class. The historical exercise should therefore be interpreted as a robustness and minimality check, not as an a posteriori significance calculation.

\subsection{Robustness to the simplicity measure}

The conclusion above should not depend on a single arbitrary definition of ``simple.'' We therefore repeat the Pareto comparison with four additional description measures,
\begin{align}
C_2&=a^2+b^2+c^2+p^2+q^2,\\
C_\infty&=\max(|a|,|b|,|c|)+\max(p,q),\\
C_E&=\sqrt{a^2+b^2+c^2}+p+q,\\
C_{\rm bit}&=\sum_{x\in\{a,b,c,p,q\}} b(x),
\end{align}
where $b(0)=0$ and $b(x)=\lfloor\log_2|x|\rfloor+1$ for nonzero integers. Table~\ref{tab:complexityrobust} reports, for each measure, the complexity of the strict $gg$ point, the number of enlarged-scan candidates no more complex than it, and its score rank inside that subset.

\begin{table}[t]
\caption{Robustness of the 2018 simplicity/Pareto statement to alternative complexity measures. In every case the strict $(2,1,2),3/4$ point is the best-scoring candidate among all enlarged-scan points no more complex than itself.}
\label{tab:complexityrobust}
\begin{ruledtabular}
\begin{tabular}{lccc}
Measure & $C_{gg}$ & Candidates with $C\le C_{gg}$ & Rank\\
\hline
$C_1=|a|+|b|+|c|+p+q$ & 12 & 7974 & 1\\
$C_2$ & 34 & 4254 & 1\\
$C_\infty$ & 6 & 4070 & 1\\
$C_E$ & 10 & 7446 & 1\\
$C_{\rm bit}$ & 10 & 19142 & 1\\
\end{tabular}
\end{ruledtabular}
\end{table}

Thus the minimality statement is not an artifact of the $L^1$-like measure in Eq.~\eqref{eq:complexity}. It remains, of course, conditional on the discrete search class itself; the scan is a controlled comparison among simple rigid rotations, not a proof that no other flavor construction can fit the data.

\section{Exact effective flavor-spurion realization}
\label{sec:mass}

The mixing construction admits a compact exact realization at the level of an effective Majorana mass tensor. Define
\begin{equation}
\STBM=\UTBM\,\mathrm{diag}(-1,0,1)\,\UTBM^T
=\frac13
\begin{pmatrix}
-2&1&-1\\
1&1&2\\
-1&2&1
\end{pmatrix}.
\label{eq:sigmatbm}
\end{equation}
Rotating this tensor with the flavor-space $gg$ rotation gives
\begin{equation}
\boxed{\Sgg=Q_{gg}\STBM Q_{gg}^T.}
\label{eq:sigmagg}
\end{equation}
By construction,
\begin{equation}
(Q_{gg}\UTBM)^T\Sgg(Q_{gg}\UTBM)
=\mathrm{diag}(-1,0,1).
\label{eq:sigdiag}
\end{equation}
Therefore the most general complex symmetric Majorana mass matrix with the same eigenframe can be written as the quadratic polynomial
\begin{equation}
\boxed{M_\nu=A I+B\Sgg+C\Sgg^2.}
\label{eq:masspoly}
\end{equation}
If the complex mass eigenvalues, including Majorana phases, are $\hat m_i$, then
\begin{equation}
\hat m_1=A-B+C,\qquad
\hat m_2=A,\qquad
\hat m_3=A+B+C,
\end{equation}
so that
\begin{align}
A&=\hat m_2,\\
B&=\frac{\hat m_3-\hat m_1}{2},\\
C&=\frac{\hat m_1+\hat m_3-2\hat m_2}{2}.
\label{eq:abc}
\end{align}
Thus the orientation of the neutrino eigenframe is contained in a single real symmetric traceless tensor $\Sgg$, while the three complex coefficients encode the masses and Majorana phases. The mixing prediction is independent of those eigenvalues. Because $\Sgg$ has three distinct eigenvalues $(-1,0,+1)$, Eq.~\eqref{eq:masspoly} is also the quadratic spectral interpolation of an arbitrary mass assignment on that fixed eigenframe. Its role here is therefore to demonstrate an exact effective eigenframe realization; the polynomial form is not by itself a dynamical explanation or an independent novelty claim.

In a weak basis where the up-type quark sector is diagonal, the charged-lepton assumption underlying Eq.~\eqref{eq:master} may be represented by
\begin{equation}
H_\ell=M_\ell M_\ell^\dagger
=\VCKM\,\mathrm{diag}(m_e^2,m_\mu^2,m_\tau^2)\,\VCKM^\dagger.
\label{eq:chargedlepton}
\end{equation}
Equations~\eqref{eq:masspoly} and \eqref{eq:chargedlepton} reproduce the complete low-energy mixing construction exactly.

This effective realization does not constitute a derivation of $\Sgg$ or of $\alpha_{gg}=-3\theta_C/4$. A continuous SO(3) family symmetry is a natural language for such real rotations and symmetric traceless tensors, and explicit SO(3)/Pati--Salam flavor models with aligned flavons and quark-related charged-lepton corrections exist in the literature \cite{King:2005bj,King:2006np}. The present paper stops short of claiming a complete renormalizable ultraviolet completion. In particular, the assumptions $U_\ell\simeq U_d$ and the exact high-scale coefficient $3/4$ remain structural hypotheses to be addressed dynamically in future work.

\section{Present oscillation status and falsifiability}
\label{sec:data}

The present comparison is made with the official NuFIT 6.1 release, based on data available in November 2025 \cite{NuFIT61,Esteban:2024eli}. For the normal-ordering analysis including the tabulated atmospheric information (IC24 with SK atmospheric data), NuFIT quotes
\begin{align}
\sin^2\theta_{12}&=0.3088^{+0.0067}_{-0.0066}, &
\theta_{12}&=33.76^{+0.42^\circ}_{-0.41^\circ},\\
\sin^2\theta_{13}&=0.02248^{+0.00055}_{-0.00059}, &
\theta_{13}&=8.62^{+0.11^\circ}_{-0.11^\circ},\\
\sin^2\theta_{23}&=0.470^{+0.017}_{-0.014}, &
\theta_{23}&=43.29^{+0.96^\circ}_{-0.79^\circ},\\
\dcp&=212^{+26^\circ}_{-36^\circ}.&&
\label{eq:nufit61current}
\end{align}
The corresponding three-sigma intervals include $41.27^\circ<\theta_{23}<49.86^\circ$ and $125^\circ<\dcp<365^\circ$ \cite{NuFIT61}. The first JUNO results are already a major component of the improved solar precision, and dedicated JUNO-era studies are beginning to discriminate among minimally modified TBM patterns \cite{Esteban:2026juno,He:2026tbm}.

For the strict model,
\begin{equation}
\sin^2\theta_{12}\simeq0.3013,\qquad
\sin^2\theta_{13}\simeq0.02162.
\end{equation}
Using the one-sided NuFIT 6.1 one-dimensional errors only as a rough diagnostic, these values lie about $1.1\sigma$ and $1.5\sigma$ below the corresponding central values. Such one-dimensional pulls should not be combined into a global statistic because the released projections are marginalized and correlated.

\begin{figure}[t]
\centering
\includegraphics[width=0.72\textwidth]{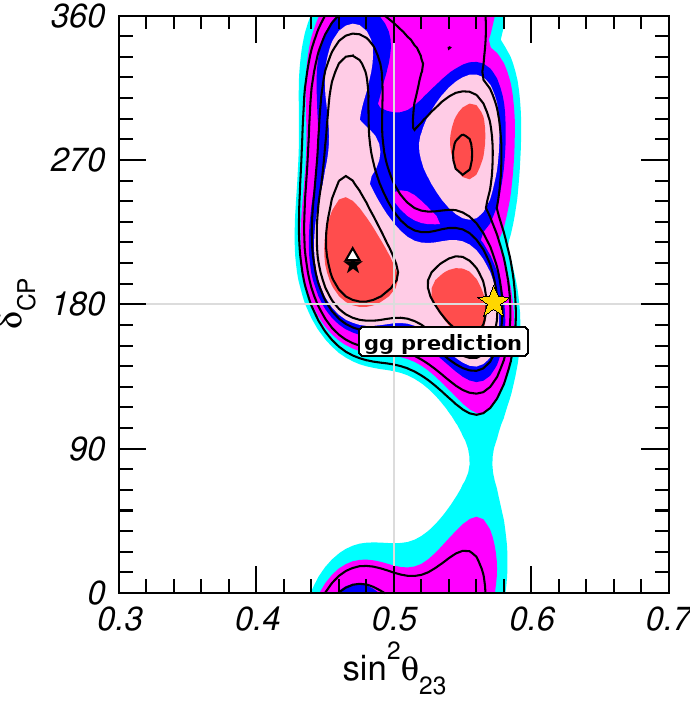}
\caption{The strict $gg$ prediction (gold star added here) overlaid on the NuFIT 6.1 normal-ordering $(\sin^2\theta_{23},\dcp)$ projection \cite{NuFIT61}. The colored regions are the NuFIT analysis without the tabulated SK/IC atmospheric $\chi^2$ information, while the black contours show the corresponding atmospheric-data-inclusive analysis; the published contours represent the standard 1$\sigma$, 90\%, 2$\sigma$, 99\% and 3$\sigma$ two-degree-of-freedom regions. The small black/white markers belong to the original NuFIT plot. The $gg$ point occupies the upper-octant, near-CP-conserving branch rather than the lower-octant global minimum.}
\label{fig:nufit2d}
\end{figure}

The atmospheric sector is the more incisive test. The $gg$ prediction is
\begin{equation}
\boxed{(\sin^2\theta_{23},\dcp)\simeq(0.5729,180.8^\circ).}
\label{eq:correlatedpoint}
\end{equation}
NuFIT 6.1 has its atmospheric-data-inclusive normal-ordering minimum in the lower octant, while retaining an upper-octant allowed branch. The strict prediction is inside the published normal-ordering allowed contours but displaced from the lower-octant global minimum, making $\theta_{23}$ the present pressure point of the construction. This correlated situation is displayed directly in Fig.~\ref{fig:nufit2d}.

We deliberately do not quote an ``exact global $\chi^2$'' obtained by adding the one-dimensional NuFIT profiles. The NuFIT 6.1 release provides marginalized one-, two- and three-dimensional $\Delta\chi^2$ tables, including a dedicated `T23/DCP' projection \cite{NuFIT61}; those projections are the appropriate public objects for correlated tests, not independent contributions to be summed. Figure~\ref{fig:nufit2d} therefore emphasizes the experimentally relevant two-dimensional comparison without assigning spurious multidimensional precision.

This makes the model sharply falsifiable. A robust future determination of the lower atmospheric octant would contradict the leading prediction in Eq.~\eqref{eq:t23exp}. Likewise, a firm observation of large leptonic CP violation with $\dcp$ far from $\pi$ would exclude the minimal real deformation, because Eq.~\eqref{eq:jcp} parametrically suppresses $\Jcp$. These failures would not be repairable by present CKM uncertainties, which induce only the narrow bands in Table~\ref{tab:pred}.

The mixing equation itself does not determine the neutrino mass ordering. In the atmospheric-data-inclusive NuFIT 6.1 analysis normal ordering is the global best fit, with inverted ordering disfavored relative to it \cite{NuFIT61}. Normal ordering is therefore the phenomenologically natural reference case for the present discussion, but it is not a mathematical prediction of Eq.~\eqref{eq:master}.

\section{Discussion}

The construction should be viewed as a constrained matrix-level realization of QLC rather than a complete flavor theory. Its economy can be summarized as follows. Once the measured CKM matrix is supplied, the strict Dirac mixing sector introduces no new continuous parameter: the remaining inputs are the discrete structural choice of the $gg$ generator, the real-deformation assumption, minimal phase alignment, and the relation $\alpha_{gg}=-3\theta_C/4$. The neutrino mass eigenvalues and Majorana phases remain unconstrained by the mixing ansatz.

\subsection{Relation to prior constructions and scope of novelty}

The individual ingredients of Eq.~\eqref{eq:master} have extensive precedents. QLC itself dates to the early 2000s \cite{Minakata:2004xt,Antusch:2005ca}; Cabibbo-sized lepton perturbations and unitary expansions about TBM are established ideas \cite{Datta:2005ci,Pakvasa:2007xv}; discrete-symmetry models have combined QLC and TBM \cite{Ahn:2011yj}; single-plane rotations of TBM and charged-lepton rotations have been used to derive mixing-angle/CP correlations \cite{Shimizu:2014tia}; and exponential parametrizations have recast quark--lepton complementarity in terms of rotation axes \cite{Zhukovsky:2019qlc}. The direct matrix-level antecedent of the present work is the numerical QLC correlation matrix reconstructed in Ref.~\cite{Sharma:2015rva}, while a particularly close modern comparator is the CKM-multiplied BM/TBM/GR analysis of Ref.~\cite{Giarnetti:2024qlc}.

Accordingly, the novelty claimed here is deliberately restricted to the \emph{combined rigid object}: the convention-covariant $gg$ generator represented by $(2,1,2)/3$ in the canonical TBM basis, the fixed angle $-3\theta_C/4$, and the resulting simultaneous zero-new-continuous-parameter prediction of the four Dirac PMNS observables once those structural hypotheses are imposed. A generic rigid rotation can be decomposed into ordinary Euler rotations, so the mathematical availability of such a matrix is not new; the distinguishing restriction is that the equivalent Euler angles are not independent parameters but are all fixed by one Cabibbo-locked axis-angle transformation. To our knowledge, the literature audit summarized above has not identified an earlier construction equivalent to this full rigid $gg$ prescription under TBM convention changes. This wording should not be read as a proof of exhaustive uniqueness.

The leading coefficient in Eq.~\eqref{eq:jcp} is also not claimed as new: it already appears in the CKM--TBM parent structure of Ref.~\cite{Giarnetti:2024qlc}. What is specific to the present model is that the same CKM-induced CP suppression coexists with the rigid $gg$ deformation and the correlated angle predictions in Eq.~\eqref{eq:predictions}.

\subsection{QLC versus gauge quark--lepton unification}

The terminology is important. Equation~\eqref{eq:master} establishes a quark--lepton mixing relation of the QLC type; it does not unify quarks and leptons into common gauge multiplets. Contemporary GUT analyses can correlate full quark and lepton Yukawa structures, neutrino masses, heavy-neutrino spectra and other observables within explicit groups such as SO(10) \cite{Chen:2026gut}. By contrast, no unified gauge group, symmetry-breaking sector, leptoquark spectrum or complete fermion-mass model is assumed here. We therefore refer to the present proposal as a \emph{minimal QLC flavor-correlation construction}, reserving ``quark--lepton unification'' for a possible ultraviolet embedding.

The retrospective scan adds a useful but limited form of evidence. Because the structure was discovered after modern data were known, the historical exercise cannot turn into a blind prediction. What it establishes is that, within a broad class of simple oriented integer axes and rational Cabibbo coefficients, the same structure would already have been exceptionally competitive using 2016--2018 information, that the 2018 continuous optimum for its rotation strength lies at $0.74941\simeq3/4$, and that the simplicity conclusion survives several alternative complexity measures. These observations support the rational choice as a compact hypothesis; they do not derive a fundamental $3/4$ law.

A complete theory would still need to explain why the physical flavor generator corresponding to Eq.~\eqref{eq:gamma} is selected, why the quark and charged-lepton left rotations are aligned as assumed, and why the common deformation is locked to the Cabibbo angle with coefficient $3/4$. Those questions are logically separable from the low-energy statement tested here and are deferred to a dedicated ultraviolet analysis.

\section{Conclusions}
\label{sec:conclusion}

We have presented a minimal, exactly unitary realization of three-family quark--lepton complementarity based on the correlation matrix
\begin{equation}
\Vgg=\UTBM R_{(2,1,2)/3}\!\left(-\frac34\theta_C\right),
\end{equation}
and the PMNS relation
\begin{equation}
\boxed{
\UPMNS=\VCKM^\dagger\UTBM
R_{(2,1,2)/3}\!\left(-\frac34\theta_C\right).
}
\end{equation}
For current CKM inputs the strict construction predicts
\begin{equation}
\boxed{
\theta_{12}\simeq33.29^\circ,\quad
\theta_{13}\simeq8.46^\circ,\quad
\theta_{23}\simeq49.19^\circ,\quad
\dcp\simeq180.79^\circ.
}
\end{equation}
The leading Wolfenstein expansion explains the upper-octant atmospheric angle and preserves the CKM--TBM leading behavior $\Jcp=-A\eta\lambda^3/6+\mathcal O(\lambda^4)$, making near-CP conservation a structural prediction of the real ansatz without claiming that coefficient itself as new.

A retrospective epoch-matched scan over 5510 simple oriented integer/rational alternatives ranks the strict $gg$ point first for both 2016 and 2018 data. In an enlarged 49,622-candidate 2018 scan it is second in raw score but first at or below its description complexity, while a continuous profile of the rotation strength gives $r_{\rm BF}=0.74941$. These results support minimality rather than absolute uniqueness.

The coordinate axis can be replaced by the convention-covariant flavor generator $\Ggg$, and the same mixing eigenframe admits the exact effective Majorana realization $M_\nu=A I+B\Sgg+C\Sgg^2$. Thus the low-energy construction can be represented by one flavor-orientation tensor while leaving neutrino masses and Majorana phases independent.

Present oscillation data do not exclude the strict model in normal ordering, but the atmospheric sector is now its main pressure point: NuFIT 6.1 places the global minimum in the lower octant while the $gg$ point occupies the allowed upper-octant, near-CP-conserving branch. The increasingly precise solar angle provides a second precision stress test. A firm lower-octant result or large leptonic CP violation away from $\pi$ would rule out the minimal real $\Vgg$ realization. The model is therefore both economical and experimentally vulnerable, which is the main motivation for treating Eq.~\eqref{eq:master} as a concrete QLC/flavor-correlation hypothesis rather than as a flexible parametrization or a completed theory of quark--lepton unification.

\begin{acknowledgments}
Acknowledgments and funding information to be inserted before submission.
\end{acknowledgments}

\appendix

\section{Exact Rodrigues matrix for the canonical $gg$ axis}
\label{app:rotation}

For $n_{gg}=(2,1,2)^T/3$, $c=\cos\alpha$ and $s=\sin\alpha$, Eq.~\eqref{eq:rodrigues} becomes
\begin{equation}
R_{gg}(\alpha)=\frac19
\begin{pmatrix}
4+5c & 2(1-c)-6s & 4(1-c)+3s\\
2(1-c)+6s & 1+8c & 2(1-c)-6s\\
4(1-c)-3s & 2(1-c)+6s & 4+5c
\end{pmatrix}.
\label{eq:exactR}
\end{equation}
The exact relations
\begin{equation}
R_{11}=R_{33},\qquad R_{12}=R_{23},\qquad R_{21}=R_{32}
\end{equation}
are direct consequences of the axis geometry.

\section{Reproducibility and statistical conventions}
\label{app:repro}

All matrix multiplications in the numerical analysis use the exact PDG standard CKM parametrization. Historical Wolfenstein inputs are converted to the exact standard-angle matrix before the PMNS parameters are extracted. The PMNS angles are obtained from matrix-element moduli in the usual standard convention, and $\dcp$ is reconstructed from both the Jarlskog invariant and a rephasing-invariant expression for $\cos\dcp$.

For a historical observable with published best fit $x_0$ and asymmetric one-sigma errors $(\sigma_-,\sigma_+)$, the scan score uses
\begin{equation}
\chi^2_{\rm score}=\sum_i
\left(\frac{x_i-x_{0,i}}{\sigma_i(x_i)}\right)^2,
\qquad
\sigma_i(x_i)=
\begin{cases}
\sigma_{+,i},&x_i\ge x_{0,i},\\
\sigma_{-,i},&x_i<x_{0,i}.
\end{cases}
\end{equation}
This statistic is used only to rank the discrete historical candidates. It is not interpreted as the full NuFIT likelihood, and confidence claims in the main text are not based on adding independently marginalized NuFIT curves.

A self-contained Python implementation of the exact ansatz, historical scans, continuous $r$ profile, and CKM-error propagation is supplied as ancillary material with this manuscript.

\bibliography{references}

\end{document}